\documentclass[a4paper,11pt]{article}
\usepackage{pos}
\usepackage{bbold}
\usepackage{lipsum}

\title{Toward tensor renormalization group study of lattice QCD}

\author*{Atis Yosprakob}

\affiliation{Department of Physics, Niigata University,\\
   Niigata 950-2181, Japan}

\emailAdd{ayosp@phys.sc.niigata-u.ac.jp}

\abstract{
The tensor renormalization group is a promising complementary approach to traditional Monte Carlo methods for lattice systems, as it is inherently free from the sign problem. We discuss recent developments crucial for its application to lattice QCD: the multi-layer construction for multi-flavor gauge theory and the armillary sphere formulation for non-Abelian gauge theory. These techniques are important for reducing the size of the initial tensor and for eliminating non-local entanglement structures within the tensor network. We present selected numerical results and discuss potential generalizations to lattice QCD.
}

\FullConference{The 41st International Symposium on Lattice Field Theory (LATTICE2024)\\
 28 July - 3 August 2024\\
Liverpool, UK\\}

\begin{document}
\maketitle

\section{Introduction}

In recent years, tensor networks have found increasing applications in high-energy physics. Perhaps the biggest motivation is the fact that tensor network algorithms are free from the sign problem by construction, making it a promising complement to the Monte Carlo methods. The tensor renormalization group (TRG) is a class of tensor network algorithms that employ a coarse-graining procedure to directly evaluate partition functions. Although originally developed for two-dimensional classical spin systems \cite{Levin:2006jai}, these algorithms have since been generalized to various systems, including scalar, gauge, and fermionic theories \cite{Kadoh:2019ube,Akiyama:2020ntf,Akiyama:2021zhf,Kuramashi:2019cgs,Fukuma:2021cni,Hirasawa:2021qvh,Kuwahara:2022ubg,Gu:2013gba,Akiyama:2024ush,Shimizu:2014uva,Shimizu:2014fsa,Shimizu:2017onf}. Efficient algorithms for higher-dimensional systems have also been proposed \cite{Xie:2012mjn,Adachi:2019paf,Kadoh:2019kqk}. In principle, we already have the necessary ingredients and methodologies for lattice QCD; however, several practical challenges remain to be addressed.

Lattice QCD is a well-established non-perturbative framework for studying strong interactions. It describes a 3+1 dimensional theory with non-Abelian gauge symmetry and multiple fermion flavors. Consequently, the local Hilbert space becomes prohibitively large for tensor network approaches due to the theory’s rich symmetries. Furthermore, it is well known that the internal symmetry within the tensor network of a non-Abelian gauge theory induces a non-local entanglement structure, manifesting as a severe degeneracy in the singular value spectrum \cite{Fukuma:2021cni}. In this article, we summarize two key concepts developed in the author’s recent work \cite{Yosprakob:2023tyr,Yosprakob:2023jgl,Yosprakob:2024sfd} to address these challenges. The first is the multi-layer construction for multi-flavor gauge theories, and the second is the armillary sphere formulation for non-Abelian gauge theories. We conclude by discussing the potential generalizations of these approaches towards lattice QCD.

\section{Tensor network representations for lattice gauge theory}

In this section, we discuss how gauge theories are typically formulated as tensor networks. Consider a gauge theory with gauge group $G$ and $N_f$ flavors of Wilson fermions labeled by $\alpha = 1, \cdots, N_f$. We assume that the fermion of flavor $\alpha$ transforms under an irreducible representation $r_\alpha$ of the gauge group. Let $U^r_{x,\mu}$ denote the link variable located at site $x$ in the direction $\hat{\mu}$, transforming in the representation $r$. If $r$ is not explicitly specified, we assume the fundamental representation, $\textbf{fund}$. We will consider the following partition function in this discussion:

\begin{align}
    Z &= \int dU d\bar\psi d\psi \exp\left\{ -S_g[U] - \sum_{x}\sum_{\alpha=1}^{N_f}\bar\psi^{(\alpha)}_x D\!\!\!\!/\;^{(\alpha)}\psi^{(\alpha)}_x\right\};\\
    S_g[U]&=\frac{\beta}{N}\sum_{1\leq\mu<\nu\leq D}\sum_{x}\mathfrak{Re}\text{Tr}\left\{\mathbb{1}-U_{x,\mu}U_{x+\hat\mu,\nu}U^\dagger_{x+\hat\nu,\mu}U^\dagger_{x,\nu}\right\},\\
    D\!\!\!\!/\;^{(\alpha)}\psi^{(\alpha)}_x&=
    -\frac{1}{2}\sum_{\nu=1}^D\left\{
    (\mathbb{1}-\gamma_\nu)e^{+\tilde\mu_\alpha\delta_{\nu D}}U^{r_\alpha}_{x,\nu}\psi^{(\alpha)}_{x+\hat\nu}
    +(\mathbb{1}+\gamma_\nu)e^{-\tilde\mu_\alpha\delta_{\nu D}}(U^{r_\alpha}_{x-\hat\nu,\nu})^\dagger\psi^{(\alpha)}_{x-\hat\nu}
    \right\}\label{eq:dirac_op}\\
    &\qquad + (\tilde m_\alpha+D)\psi^{(\alpha)}_x.\nonumber
\end{align}

The additional parameters are the inverse coupling $\beta$, the `group size' $N:=\text{dim}(\textbf{fund})$, the dimensionless chemical potential $\tilde{\mu}_\alpha$, and the dimensionless mass $\tilde{m}_\alpha$. The direction $\mu = D$ is taken to be the imaginary time direction, and $\gamma_\mu$ denotes the Gamma matrices.
It is convenient to separate the fermion bilinears into on-site terms and hopping terms.
\begin{align}
    \bar\psi^{(\alpha)}_x D\!\!\!\!/\;^{(\alpha)}\psi^{(\alpha)}_x &=
    \bar\psi^{(\alpha)}_x W_x^{(\alpha)}\psi^{(\alpha)}_x
    +\sum_\nu\left\{\bar\psi^{(\alpha)}_x H_{x,\nu}^{(\alpha)}\psi^{(\alpha)}_{x+\hat\nu}
    +\bar\psi^{(\alpha)}_x H_{x,-\nu}^{(\alpha)}\psi^{(\alpha)}_{x-\hat\nu}\right\};\\
    W^{(\alpha)}_x &:= \tilde m_\alpha+2,\\
    H^{(\alpha)}_{x,\nu}&=-\frac{1}{2}(\mathbb{1}-\gamma_\nu)e^{+\tilde\mu_\alpha\delta_{\nu D}}U^{r_\alpha}_{x,\nu},\label{eq:Hplus}\\
    H^{(\alpha)}_{x,-\nu}&=-\frac{1}{2}(\mathbb{1}+\gamma_\nu)e^{-\tilde\mu_\alpha\delta_{\nu D}}(U^{r_\alpha}_{x-\hat\nu,\nu})^\dagger\label{eq:Hminus}.
\end{align}

\begin{figure}
    \centering
    \includegraphics[width=0.45\linewidth]{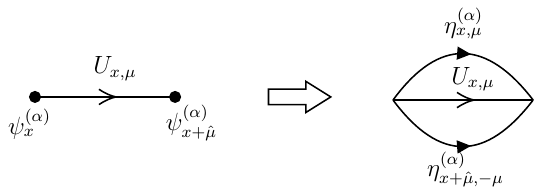}
    \caption{The transformation of site fermions $\psi$ into link fermions $\eta$. The gauge link variable $U$ is shown for reference.}
    \label{fig:site_link_transform}
\end{figure}

To make this action suitable for the tensor network construction, we transform the site fermions $\psi^{(\alpha)}_x$ into link fermions $\eta^{(\alpha)}_{x,\mu}$, as described in Ref.~\cite{Akiyama:2020sfo}, resulting in the partition function \cite{Yosprakob:2023tyr}:

\begin{align}
    Z &= \int dU \int_{\bar\eta\eta} e^{-S_g[U]}\prod_x \mathcal{F}_x;\\
    \mathcal{F}_x &= \prod_{\alpha}\int d\bar\psi_x^{(\alpha)} d\psi_x^{(\alpha)} e^{-\bar\psi^{(\alpha)}_x W_x^{(\alpha)}\psi^{(\alpha)}_x-{\displaystyle\sum_{\pm,\nu}}\left\{\bar\psi_x^{(\alpha)}\eta_{x,\pm\nu}^{(\alpha)}-\bar\eta_{x\mp\hat\nu,\pm\nu}^{(\alpha)}H_{x\mp\hat\nu,\pm\nu}^{(\alpha)}\psi_{x}^{(\alpha)}\right\}},
\end{align}
The fermionic integral $\mathcal{F}_x$ can be evaluated analytically, yielding a polynomial function of the link variables $U$ and $\eta$ with a finite number of terms.
The symbol $\int_{\bar\eta\eta}$ is a shorthand notation for
\begin{equation}
    \int_{\bar\eta\eta} := \int \prod_{x,\nu,\alpha} d\bar\eta^{(\alpha)}_{x,\nu} d\eta^{(\alpha)}_{x,\nu} e^{-\bar\eta^{(\alpha)}_{x,\nu} \eta^{(\alpha)}_{x,\nu}},
\end{equation}
which represents the measure for the fermionic tensor contraction. The site-link transformation is illustrated in Figure \ref{fig:site_link_transform}.

Since the gauge fields are already placed on the links, we can use the Haar measure $\int dU$ for the bosonic contraction. For the tensor networks, this can be implemented in various ways, such as using quadrature \cite{Kuramashi:2019cgs,Yosprakob:2023tyr} or ensemble sampling \cite{Fukuma:2021cni,Kuwahara:2022ubg}. The resulting partition function as a tensor network is

\begin{align}
    Z &= \int_{\bar\eta\eta} \int dU \prod_{x} \left( \mathcal{F}_x \times \prod_{\mu<\nu} P_{x,\mu\nu} \right); \label{eq:direct_TN} \\
    P_{x,\mu\nu} &= \exp \left[ \frac{\beta}{N} \mathfrak{Re} \text{Tr} \left\{ U_{x,\mu} U_{x+\hat\mu,\nu} U^\dagger_{x+\hat\nu,\mu} U^\dagger_{x,\nu} - \mathbb{1} \right\} \right]. \label{eq:plaquette_tensor}
\end{align}

\begin{figure}
    \centering
    \includegraphics[scale=0.6]{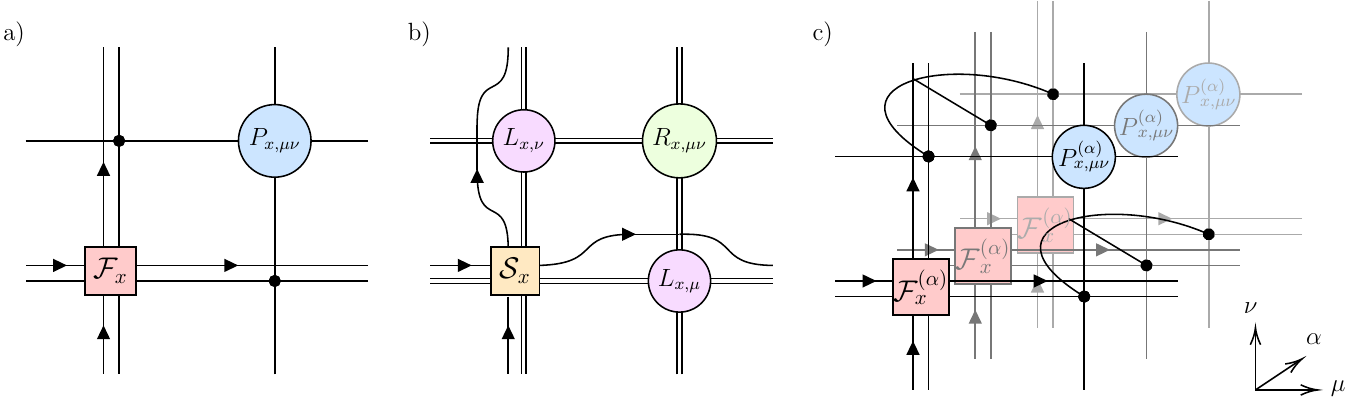}
    \caption{The tensor connections in a unit cell on site $x$; a) with the direct group integral, b) with the character expansion, and c) the multi-layer construction of a). Single lines without arrows correspond to bosonic link $U$, single lines with arrows correspond to fermionic links $\eta$, and double lines correspond to the representation links $(r,i,j)$. The black nodes are delta functions ensuring that all incoming links take the same value. This diagram applies for any $\mu\!-\!\nu$ plane in $D$ dimensions.}
    \label{fig:unit_cell}
\end{figure}

Here, $P_{x,\mu\nu}$ is a tensor located on the plaquette, with four legs corresponding to the four link variables in \eqref{eq:plaquette_tensor}, while $\mathcal{F}_x$ is a Grassmann tensor located on the site with $2D$ bosonic legs (the link variables in \eqref{eq:Hplus} and \eqref{eq:Hminus}) and $2D$ fermionic legs of dimension $2^{2N_f}$ (there are two link fermions for each flavor; see Figure \ref{fig:site_link_transform}). The diagrammatic representation of \eqref{eq:direct_TN} is shown in Figure \ref{fig:unit_cell}-a).

In the case where $ G $ is a non-Abelian gauge group, it is more informative to use the character expansion on $ P_{x,\mu\nu} $ and consider the representation sum as a tensor contraction, rather than as a group integral \cite{Liu:2013nsa,Hirasawa:2021qvh,Yosprakob:2023jgl}. Consider the character expansion:
\begin{equation}
    P_{x,\mu\nu} = \sum_{r \in \text{irreps}} f_r(\beta/N) \, \text{Tr}\left\{ U^r_{x,\mu} U^r_{x+\hat{\mu},\nu} \left( U^r_{x+\hat{\nu},\mu} \right)^\dagger \left( U^r_{x,\nu} \right)^\dagger \right\},
\end{equation}
with the coefficient
\begin{equation}
    f_r(\beta/N) = \int dU \, \text{Tr}\left( U^r \right)^\dagger e^{\frac{\beta}{N} \mathfrak{Re} \, \text{Tr}\{ U - \mathbb{1} \}}.
\end{equation}
We can also perform a series expansion on $ \mathcal{F}_x $:
\begin{equation}
    \mathcal{F}_x = \sum_{r_1, \dots, s_D \in \{\textbf{trv},\textbf{fund}\}} \sum_{\{i_\mu,j_\mu,k_\mu,l_\mu\}} (\mathcal{S}_x)^{r_1 \dots s_D}_{i_1 j_1 \dots k_D \ell_D} \prod_{\mu=1}^D \left( U_{x - \hat{\mu}, \mu}^{r_\mu} \right)_{i_\mu j_\mu} \left( U_{x, \mu}^{s_\mu} \right)^\dagger_{k_\mu \ell_\mu}.
\end{equation}
Here, the representations sum only over the trivial and fundamental representations because, for each term in $ \mathcal{F}_x $, a link variable either appears as a fundamental representation or does not appear at all (equivalently as the trivial representation). The expansion coefficient $ \mathcal{S}_x $ is a Grassmann tensor with the same fermionic legs as $ \mathcal{F}_x $, but the bosonic legs now correspond to the representation and matrix indices; $ (r_\mu,i_\mu,j_\mu) $ and $ (s_\mu,k_\mu,\ell_\mu) $.

We can group the same bosonic link variables together and perform the integral directly:
\begin{equation}
    L_{x,\mu}
    =\int dU
    \underset{\text{from }P_{x,\mu\nu}}{
    \underbrace{
    \prod_{\nu\ne\mu}(U_{x,\mu}^{r_\nu})_{i_\nu j_\nu}(U_{x,\mu}^{s_\nu})^\dagger_{k_\nu \ell_\nu}
    }}
    \times
    \underset{\text{from }\mathcal{F}_{x+\hat\mu}}{
    \underbrace{
    (U_{x,\mu}^r)_{i_\mu j_\mu }
    }}
    \times
    \underset{\text{from }\mathcal{F}_{x}}{
    \underbrace{
    (U_{x,\mu}^s)^\dagger_{k_\mu \ell_\mu }
    }}
\end{equation}
This gives the partition function:
\begin{align}
    Z &= \int_{\bar\eta\eta}\sum_{\{(r,i,s)\}}\prod_x\left(\mathcal{S}_x\prod_\mu L_{x,\mu}\prod_{\mu<\nu}R_{x,\mu\nu}\right);
    \label{eq:CE_TN}\\
    R_{x,\mu\nu}&=f_{r_1}(\beta/N)
    \delta_{r_1,r_2,r_3,r_4}
    \delta_{j_1i_2}\delta_{j_2i_3}\delta_{j_3i_4}\delta_{j_4i_1}.\label{eq:R-tensor}
\end{align}
The diagrammatic representation of \eqref{eq:CE_TN} is shown in Figure \ref{fig:unit_cell}-b).

\section{Multi-layer construction for multi-flavor gauge theories}

As described in the previous section, both $\mathcal{F}_x$ and $\mathcal{S}_x$ have $2D$ fermionic legs, each with dimension $2^{2N_f}$. This causes the total number of components for both tensors to grow exponentially with the number of flavors. Such a prohibitive cost can be avoided by either using flavor staggering \cite{Asaduzzaman:2023pyz, Pai:2024tip, Kanno:2024elz} or extending the flavors to extra dimensions \cite{Akiyama:2023lvr}. We consider the second option since we can directly apply it to our Wilson fermions without additional considerations.

For demonstrative purposes, we consider the 2D $\mathbb{Z}_K$ gauge theory with $N_f$ flavors of Wilson fermions. The representation $r_\alpha$ in \eqref{eq:dirac_op} is now replaced by the charge $q_\alpha\in\mathbb{Z}$ of flavor $\alpha$. To extend the flavors into the extra dimension, we consider auxiliary copies of the gauge link variables $U^{(\alpha)}_{x,\mu\nu}$ and define an action local to flavor $\alpha$:
\begin{equation}
    S^{(\alpha)}=\frac{1}{N_f}S_g[U^{(\alpha)}]+\sum_{x}\bar\psi^{(\alpha)}_x D\!\!\!\!/\;^{(\alpha)}\psi^{(\alpha)}_x.
\end{equation}
The partition function now becomes
\begin{equation}
    Z = \int dU  \prod_{\alpha=1}^{N_f}\int dU^{(\alpha)}d\bar\psi^{(\alpha)} d\psi^{(\alpha)}\delta(U^{(\alpha)}-U)e^{-S^{(\alpha)}}
\end{equation}
We can then proceed to construct the tensor network for each layer $\alpha$ and arrive at \eqref{eq:direct_TN}. The only differences are that the black nodes in Figure \ref{fig:unit_cell}-a), corresponding to the Kronecker deltas, now have an extra leg extending to the extra dimension, and both $\mathcal{F}_x$ and $P_{x,\mu\nu}$ are now flavor-specific. The tensor network in a unit cell is shown in Figure \ref{fig:unit_cell}-c).

\begin{figure}
    \centering
    \includegraphics[scale=0.5]{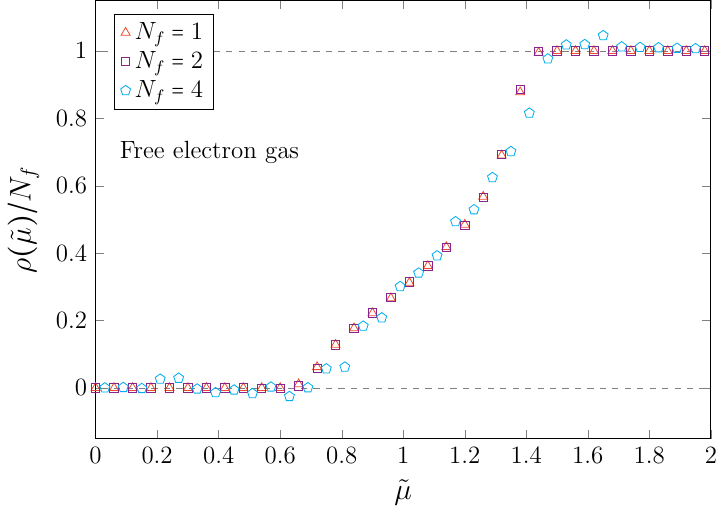}
    \includegraphics[scale=0.5]{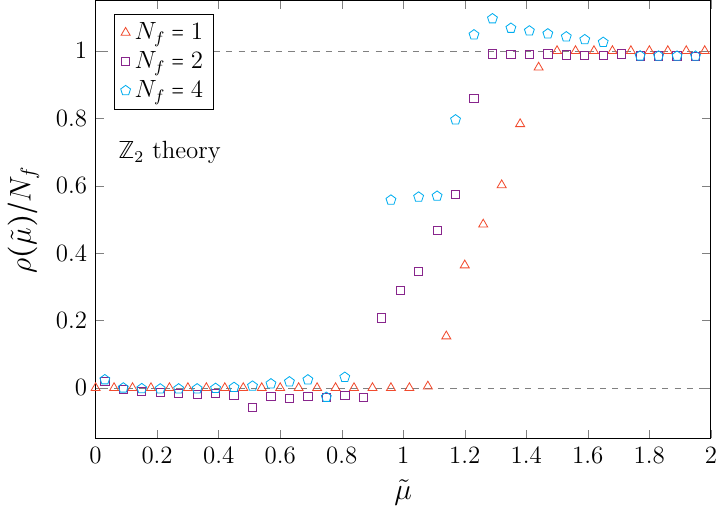}
    \caption{The number density $\rho=\frac{1}{V}\frac{d}{d\tilde\mu}\log Z$ per flavor as a function of $\tilde\mu$, with $N_f=1$, $2$, and $4$ for (left) free electron gas and (right) $\mathbb{Z}_2$ gauge theory.}
    \label{fig:finite_density}
\end{figure}

Next, we present selected results from Ref.~\cite{Yosprakob:2023tyr}. In our calculation, we first perform the compression of the tensors in a unit cell into a single tensor using higher-order SVD (HOSVD) \cite{doi:10.1137/S0895479896305696}. Due to the sparseness of the Grassmann tensor, we were able to perform the compression with a compression ratio between $10^{-4}$ and $10^{-9}$. It should be noted that the interactions between layers are still considered all-to-all (because all flavors experience the same gauge field), even though the tensor network is written in a local form. This makes the entanglement along the flavor direction very strong. As such, we first perform the coarse-graining procedure in the flavor direction using higher-order TRG (HOTRG) \cite{Xie:2012mjn}, and then perform the coarse-graining procedure in the space-time plane using the Levin-Nave TRG \cite{Levin:2006jai}. All of the Grassmann tensor network computation are done using a Python package \verb|GrassmannTN|~\cite{Yosprakob:2023flr}. In the finite-density results shown in Figure \ref{fig:finite_density}, we use $\chi_\text{HOTRG}=64$ for $N_f=1,2$, $\chi_\text{HOTRG}=32$ for $N_f=4$, and $\chi_\text{TRG}=64$ for all $N_f$. The computations are done with a volume of $V=128^2$. The Silver Blaze phenomenon, a feature that is difficult to obtain in typical Monte Carlo methods due to the sign problem, can be clearly seen.

\section{The armillary sphere formulation for non-Abelian gauge theories}

It was shown in Ref.~\cite{Fukuma:2021cni} that pure Yang-Mills theory suffers from a severe degeneracy of singular values due to non-local entanglement structures in the tensor network. In two dimensions, using character expansion, this can be easily identified as the matrix index loop around each site. Once these loops are contracted, the TRG analysis can be performed with significantly higher accuracy for the same bond dimensions \cite{Hirasawa:2021qvh}. In higher dimensions, the degeneracy is expected to become more severe. Fortunately, it is also possible to eliminate these non-local entanglement structures using the armillary sphere formalism \cite{Yosprakob:2023jgl}, which will be discussed below.

\begin{figure}
    \centering
    \includegraphics[scale=0.6]{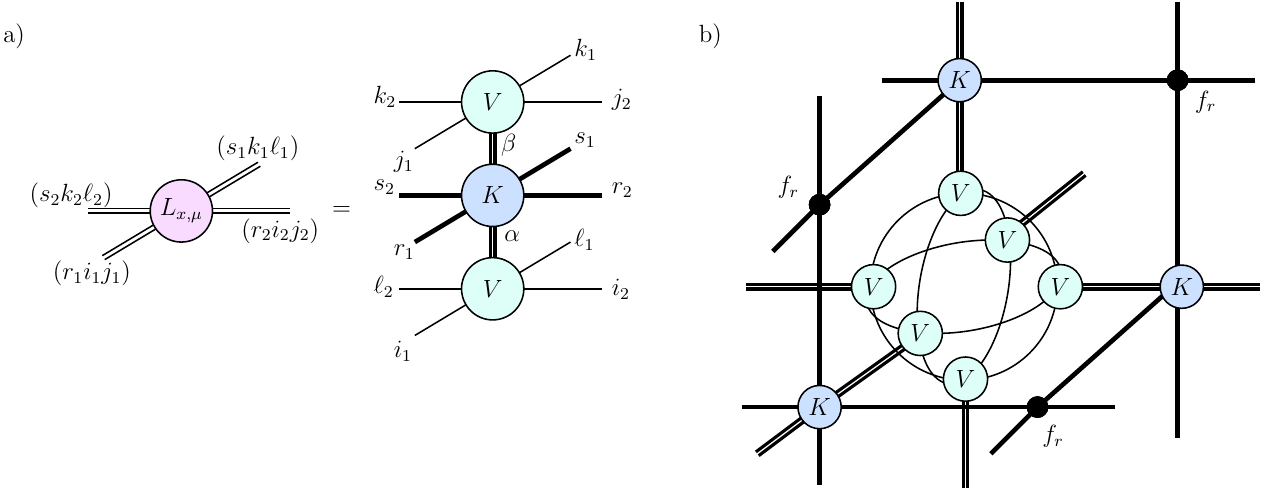}
    \caption{a) The link tensor $L_{x,\mu}$ expressed in terms of vertex tensors $V$ and a link-conditioned tensor $K$. b) A three-dimensional example of a tensor network for a pure gauge theory, represented in terms of $V$ and $K$. An armillary sphere is shown around the lattice site, with $V$ as its vertices.}
    \label{fig:armillary}
\end{figure}

For simplicity, we will consider a pure SU($N$) Yang-Mills theory. In this case, the link tensor in \eqref{eq:CE_TN} only includes the contributions from the adjacent plaquettes:
\begin{equation}
    L_{x,\mu}
    =\int dU
    \prod_{\nu\ne\mu}(U_{x,\mu}^{r_\nu})_{i_\nu j_\nu}(U_{x,\mu}^{s_\nu})^\dagger_{k_\nu \ell_\nu}.\label{eq:VKV_integral_pre}
\end{equation}
Using the Clebsch-Gordan decomposition and the grand orthogonality relation, the integral can be computed analytically \cite{Yosprakob:2023jgl,Yosprakob:2024sfd}:
\begin{equation}
    L_{x,\mu}=\sum_{\alpha,\beta}K^{\alpha\beta}_{\{r_\nu\}\{s_\nu\}}V^\alpha_{\{i_\nu\}\{\ell_\nu\}}V^\beta_{\{j_\nu\}\{k_\nu\}},\label{eq:VKV_integral}
\end{equation}
where $V$ is a `vertex tensor' composed of Clebsch-Gordan coefficients, and $K$ is the `link-conditional tensor' ensuring consistency among the representation indices. The indices $\alpha$ and $\beta$ are the `multi-representation' indices, which encode how the irreps are coupled. A full description of the formulation is given in Ref.~\cite{Yosprakob:2023jgl}. The three-dimensional example of $L_{x,\mu}$ is shown diagrammatically in Figure \ref{fig:armillary}-a). After performing the integral for every link, the vertex tensors form a closed structure around each site, which we call the `armillary sphere', named after an ancient astronomical device used to represent the great circles of the heavens. The three-dimensional example of the unit cell tensor network in terms of $V$ and $K$ is shown in Figure \ref{fig:armillary}-b). In this form, the matrix indices, which carry the non-local entanglement structures, can be traced out analytically, leaving only the harmless representation indices. 

\begin{figure}
    \centering
    \includegraphics[scale=0.5]{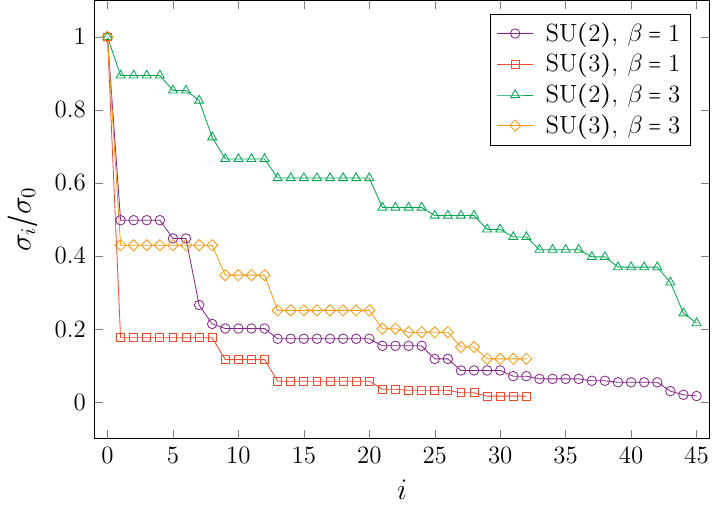}
    \caption{The singular value spectra of SU(2) and SU(3) gauge theories obtained from the HOSVD of the initial tensor without truncation. No severe degeneracy of singular values associated with the internal symmetry is observed here.}
    \label{fig:spectrum}
\end{figure}

Note that we originally have $R_{x,\mu\nu}$ \eqref{eq:R-tensor} on the plaquette center. However, since the matrix indices are grouped together in the armillary sphere, we are only left with the expansion coefficients $f_r$ and the diagonal Kronecker deltas (black nodes). Singular value spectra of SU(2) and SU(3) gauge theories with $\beta=1$ and $3$ are shown in Figure \ref{fig:spectrum}. We do not observe the degeneracy associated with the internal symmetry, i.e., the non-local entanglement structures, which tend to become increasingly severe toward the tail of the spectrum.

\begin{figure}
    \centering
    \includegraphics[scale=0.5]{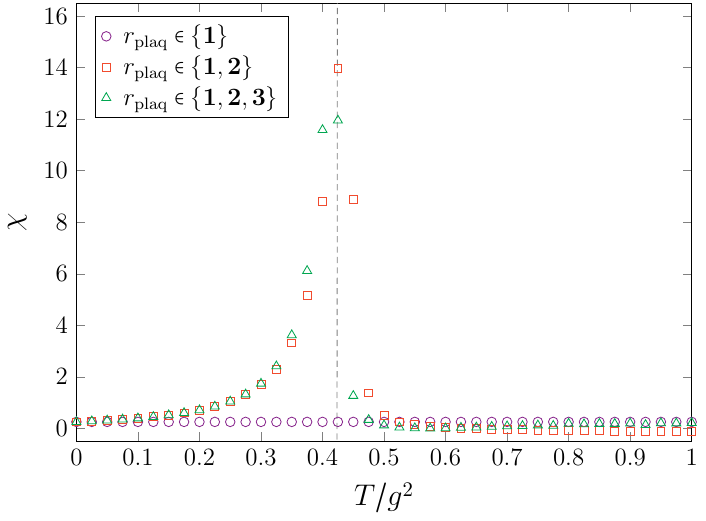}
    \includegraphics[scale=0.5]{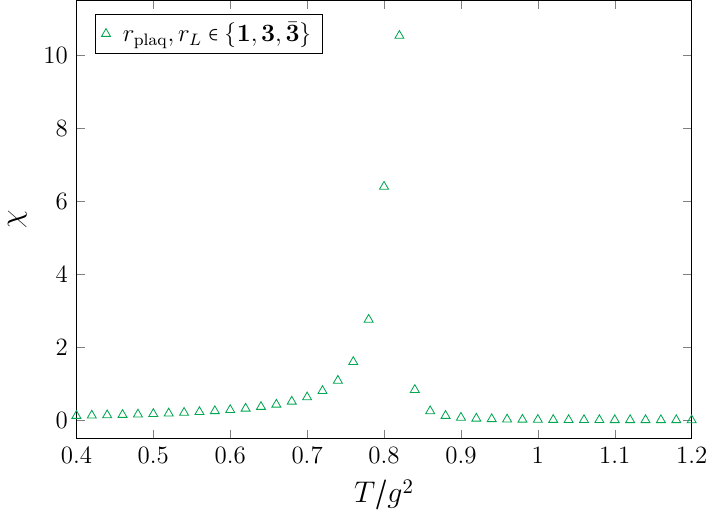}
    \caption{Polyakov susceptibility as a function of temperature $T/g^2 = \beta/2NN_\tau$ for SU(2) (left) and SU(3) (right) gauge theories. Different symbols represent the data with different numbers of terms in the character expansions: $r_\text{plaq}$ corresponds to the representations from the plaquette action, and $r_L$ corresponds to the representations from the Polyakov loop term. For SU(2), we take $r_L \in \{\textbf{1},\textbf{2}\}$. The dashed line indicates the transition temperature obtained from the Monte Carlo simulation.}
    \label{fig:polyakov}
\end{figure}

In the following, we present the results for the finite-temperature three-dimensional SU(2) and SU(3) gauge theories with a Polyakov loop source term:
\begin{equation}
    S_L[U] = \frac{\kappa}{N}\sum_{x_2\in\Lambda_2}\mathfrak{Re}\text{Tr}\left(\mathbb{1}-\prod_{\tau=1}^{N_\tau}U_{x_2+\tau\hat 3,3}\right).
\end{equation}
To access a sufficiently high temperature with a small $\beta$, we consider only one temporal time slice\footnote{For $N_\tau \geq 2$, we need to use larger $\beta$ values to reach the deconfined phase. This is challenging in the current setup, as the character expansion becomes less accurate at large $\beta$.}. The Polyakov loop susceptibility is computed using the second numerical derivative with respect to $\kappa$ at $\kappa=0.01$ and $\Delta\kappa=0.01$. The computation performed at $\kappa > 0$ ensures the spontaneous breaking of the $\mathbb{Z}_N$ center symmetry in the deconfined phase. The coarse-graining procedure on the $xy$ plane is carried out using the Levin-Nave TRG with $D_\text{cut}=96$. The results are shown in Figure \ref{fig:polyakov}. The deconfinement transition is observed for both theories. The deconfinement temperature for SU(2) is consistent with the Monte Carlo result \cite{Teper:1993gp}, confirming that the armillary sphere formulation can capture essential topological features of the gauge theory with as few as three expansion terms.

\section{Summary and outlooks}
We discuss two improvements to the tensor network formulations of lattice gauge theories for their application to lattice QCD. These are the multi-layer construction and the armillary sphere formulation, which are used to respectively reduce the size of the initial tensor for multi-flavor gauge theories and non-Abelian gauge theories. Additionally, the armillary sphere formulation can be used to eliminate non-local entanglement structures in the tensor network, which is a significant obstacle in numerical simulations. We demonstrate the two techniques with two-dimensional $N_f$-flavor $\mathbb{Z}_N$ gauge theories at finite density, as well as three-dimensional SU(2) and SU(3) pure gauge theories.

The armillary sphere formulation can be generalized straightforwardly to any pure gauge action, provided that the gauge is not fixed. (If the gauge is fixed, as is typically done in gauge-Higgs models, the action is no longer a class function of the gauge group, and the character expansion is no longer available.) This includes more complicated terms such as the improved action or the theta term. In such cases, one only needs to perform the character expansion for each term separately. By including fermions, the armillary tensor can still be constructed, but some matrix indices will remain connecting the armillary tensor and the fermionic tensor $\mathcal{S}_x$. This should not pose a problem, as we can also perform the contraction of this bond analytically. In principle, the multi-layer construction can be incorporated with the armillary sphere formulation, as long as an efficient representation of the Kronecker delta (black nodes in Figure \ref{fig:unit_cell}) is provided.

\section*{Acknowledgments}
This work is supported by a Grant-in-Aid for Transformative Research Areas “The Natural Laws of Extreme Universe—A New Paradigm for Spacetime and Matter from Quantum Information” (KAKENHI Grant No. JP21H05191) from JSPS of Japan.


\bibliographystyle{JHEP}
\bibliography{ref}

\end{document}